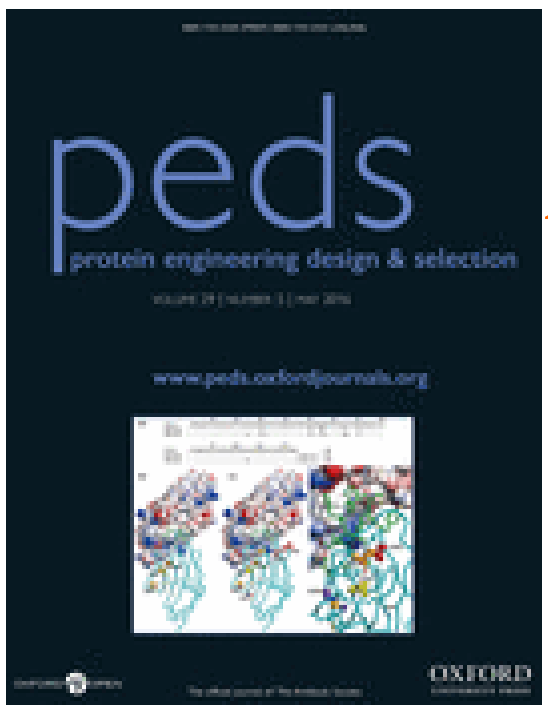

# A novel method for predicting transmembrane segments in proteins based on a statistical analysis of the SwissProt database: the PRED-TMR algorithm

**Pasquier, C., Promponas, V.J., Palaios, G.A., Hamodrakas, J.S. and Hamodrakas, S.J.**

# A novel method for predicting transmembrane segments in proteins based on a statistical analysis of the SwissProt database: the PRED-TMR algorithm

**Pasquier, C., Promponas, V.J., Palaios, G.A., Hamodrakas, J.S. and Hamodrakas, S.J.**

*Faculty of Biology, Department of Cell Biology and Biophysics, University of Athens, Panepistimiopolis, Athens 15701, Greece*

## Abstract

We present a novel method that predicts transmembrane domains in proteins using solely information contained in the sequence itself. The PRED-TMR algorithm described in this work refines a standard hydrophobicity analysis with a detection of potential termini ("edges", starts and ends) of transmembrane regions. This allows both to discard highly hydrophobic regions not delimited by clear start and end configurations and to confirm putative transmembrane segments not distinguishable by their hydrophobic composition.

The accuracy obtained on a test set of 101 non homologous transmembrane proteins with reliable topologies compares well with that of other popular existing methods. Only a slight decrease in prediction accuracy was observed when the algorithm was applied to all transmembrane proteins of the SwissProt database (release 35).

A WWW server running the PRED-TMR algorithm is available at http://o2.db.uoa.gr/PRED-TMR/

## Introduction

The prediction of protein structure is still an open problem in molecular biology. Important efforts were especially devoted to transmembrane proteins because they are involved in a broad range of processes and functions and, unfortunately, it is very difficult to solve their three-dimensional structure by X-ray crystallography (Persson and Argos, 1994; Aloy et al., 1997). For this class of proteins, structure prediction methods are needed more urgently than for globular water-soluble proteins.

A number of methods or algorithms designed to locate the transmembrane regions of membrane proteins have been developed (von Heijne, 1992; Persson and Argos, 1994; Cserzo et al., 1997). Apparently, in several cases, better results are obtained, when extra information coming from multiple alignments of homologous proteins is used (Persson and Argos, 1994; Rost et al., 1994). However, when homologies cannot be found in the databases, improvement of prediction methods using information contained in a protein sequence alone is important.

Prediction methods based on a hydrophobicity analysis can highlight most of the transmembrane regions of a protein (von Heijne, 1992). However, they fail to discriminate perfectly between segments corresponding to real transmembrane parts and simple, highly hydrophobic stretches of residues.

The algorithm presented in this paper refines information given by a hydrophobicity analysis, with a detection of favourable patterns that highlight potential termini (starts and ends) of transmembrane regions. Thus, highly hydrophobic stretches of residues that are not delimited by clear start and end configurations can be discarded. On the contrary, favourable patterns can fish out some transmembrane regions not clearly distinguishable by their hydrophobic composition.

## Methods

The aim of a prediction method is to obtain good accuracy when applied to unknown proteins. As underlined by Rost and Sander (1998), on the basis of two CASP experiments, this objective has not been reached yet. Over-optimistic results of many algorithms are usually due to the use of too small or non-representative data sets.

The PRED-TMR method, presented in this work, is based on a statistical study of transmembrane proteins. Despite the lack of precision and fidelity of SwissProt (Cserzo et al., 1997), we have chosen to collect the information needed from the whole database instead of using a limited set that may not be statistically representative.

Our method was optimised on a subset of 64 reliable proteins previously used in several prediction programs (Jones et al. 1994, Rost et al. 1995, Aloy et al. 1997) that were available in the public databases (the sequences used and the results obtained are presented on our web site at http://o2.db.uoa.gr/PRED-TMR/Results/). We relied on transmembrane segment topologies indicated in SwissProt release 35 or, when unavailable, in the paper of Rost et al., 1996.

The reliability of predictions was tested on several sets of sequences used for the rating of recent published algorithms. The PRED-TMR algorithm was also applied to the whole SwissProt database.

*Information gathering*

9392 transmembrane proteins were automatically extracted from the SwissProt database, release 35, based on the presence in the feature table of the 'TRANSMEM' keyword. The information relative to the transmembrane regions and their peripheral residues were stored in a database called DB-TMR. This database contains for each transmembrane segment:

- the access code of the sequence containing the segment (ID line),
- the organism classification (OC lines),
- the length of the transmembrane region,
- the direction of the transmembrane segment when it can be deduced from the keywords 'CYTOPLASMIC' and 'EXTRACELLULAR' of the feature table,
- Five amino-acid residues (one letter code) outside the transmembrane region for the N and C-terminal sides respectively,
- the amino-acid residues (one letter code) of the transmembrane segment.

This information can easily be filtered by organism or transmembrane type in order to refine the statistical analysis. The database and the description of the format used can be downloaded from our web site at http://o2.db.uoa.gr/DB-TMR/.

To minimise the impact of erroneous information, transmembrane segments that extend beyond the end(s) of the sequenced region or with unknown endpoints are discarded before the statistical calculations.

*Distribution of transmembrane segment length*

The 40548 transmembrane segments with reliable endpoints contained in DB-TMR have an average length of 21.30 residues and a standard deviation of 2.56 residues. The distribution is sharper than a gaussian distribution, with 60% of the transmembrane segments having a length of 21 residues and 94% having a length between 17 and 25 residues. A simple approximation of the curve is given by the function:

$$f(x) = e^{|l-21}$$

where $l$ is the length of the transmembrane segment.

*Calculation of amino acid residue transmembrane propensities (potentials)*

A propensity for each residue to be in a transmembrane region was calculated using the formula:

$$Pi = \frac{Fi^{TM}}{Fi}$$

where $Pi$ is the propensity value (transmembrane potential) of residue type $i$ and $Fi^{TM}$ and $Fi$ are the frequencies of the ith type of residue in transmembrane segments and in the entire SwissProt database respectively. Values above 1 indicate a preference for a residue to be in the lipid-associated structure of a transmembrane protein, whereas propensities below 1 characterise unfavourable transmembrane residues. The propensity values for the 20 amino acid residue are given on the web page http://o2.db.uoa.gr/PRED-TMR/material.html.

*Evaluation of the "hydrophobicity" of a sequence of residues*

Following a similar, but not identical, definition put forward by Sipos and von Heijne (1993), the table of transmembrane propensities was translated into a new, statistically based, "hydrophobicity" scale defined by:

$$Hi = \ln(Pi)$$

where $Hi$ is a measurement of the "hydrophobicity" of a residue of type $i$.

The "hydrophobicity" of a sequence of residues from position $m$ to position $p$ is evaluated by:

$$H_m^p = \sum_{k=m}^{k=p} H(R_k)$$

where $H_m^p$ is the score of the considered segment and $R_k$, the type of residue located at position k in the sequence.

*Calculation of favourable terminal (end) configurations of transmembrane regions*

Favourable configurations are computed for decapeptides centred at the border of transmembrane regions (5 residues outside and 5 residues inside the membrane). Positions in the decapeptide are counted from 0 to 9. For the N-terminal end (side), thereafter also referred to as "left end", position 0 corresponds to a residue 5 residues before the first amino acid residue of the transmembrane segment and position 9 corresponds to a residue 4 residues after this residue. For the C-terminal end (side), thereafter also referred to as "right end", position 0 corresponds to a residue 5 residues after the last amino acid residue of the transmembrane segment and position 9 corresponds to a residue 4 residues before this residue (Figure 1).

```
          ——Transmembrane segment——
01234 56789                   98765 43210
IIRRR PLFYAVSLLLPSIFLMVVDIVGFCLPP DSGERVSFKITLLLGYSVFLIIVSDTLP
|                                                              |
241                                                          300
```

**Fig. 1:** The sequence of the protein 5HT3_MOUSE (SwissProt protein code) from residue 241 to residue 300. A putative transmembrane segment as defined in the SwissProt database (release 35) is shown in grey. Digits above the sequence, which is shown in the one-letter code, indicate the nominal positions in the decapeptide(see Text) of the corresponding residues, at the N- and C-terminal ends of the transmembrane segment.

The propensity for an amino acid of type *i* to appear at position *p* in the decapeptide is defined by the formula:

$$P_i^p = \frac{F_i^p}{F_i}$$

where $P_i^p$ is the propensity value of residue type $i$ at position $p$, $F_i^p$ and $F_i$ are the frequency of the *ith* type residue at position *p* in the decapeptide and in the entire SwissProt database respectively. Clearly, values above 1 indicate a preference for the residue considered to be present at the specified position, whereas values below 1 suggest that these residues are not favoured at this position. The table of propensities for each amino acid in the decapeptide is given on the web page http://o2.db.uoa.gr/PRED-TMR/material.html.

For the N-terminal ("left") side of a transmembrane segment, the propensity $P_p^{left}$ of an amino-acid residue, at position p in the sequence, to be the first one in the lipid-associated structure (the first residue of the transmembrane domain) is defined by the formula:

$$P_p^{left} = \sum_{k=-5}^{k=4} \ln(P_{R(p+k)}^{5+k})$$

The summation is performed for the entire decapeptide, from position p-5 to position p+4.

Similarly, for the C-terminal side ("right") of a transmembrane segment, the propensity for an amino acid at position p to be the first residue outside the transmembrane region is defined by:

$$P_p^{right} = \sum_{k=-5}^{k=4} \ln(P_{R(p+k)}^{4-k})$$

Values above 0 indicate favourable configurations whereas values below 0 suggest unfavourable ones.

However, using only $P^{left}$ propensities to find good "left" configurations (or $P^{right}$ to find "right" configurations) is not sufficient. Some decapeptides can indeed generate high scores for both "left" and "right" propensities. We have, for example, to discard decapeptides like 'ILFVSTFFTM' which give a good value for $P^{left}$ of 1.75 and a high value for $P^{right}$ of 2.61.

By looking at the $P^{left}$ and $P^{right}$ values for known transmembrane segments, we found that the scores themselves are less important than the difference between "left" and "right" values.

We combined both propensities to obtain start and end indicators of transmembrane segments using the formulae:

$$LeftInd_p = \frac{P_p^{left} + \min\left(P_p^{left}, P_p^{left} - P_p^{right}\right)}{2}$$

$$RightInd_p = \frac{P_p^{right} + \min\left(P_p^{right}, P_p^{right} - P_p^{left}\right)}{2}$$

where $LeftInd_p$ is an indicator for the decapeptide centred at position $p$ to represent a start configuration of a transmembrane region and $RightInd_p$ an indicator for the same decapeptide to represent an end configuration. The minimum is used to avoid that a small $P^{right}$ contributes more than $P^{left}$ in the evaluation of the start configuration (the inverse is also true for end configurations).

*Scoring of transmembrane regions*

A well defined transmembrane region should give good scores for all three parameters (*LeftInd*, *RightInd* and *H*). However, when applied to known transmembrane segments, a large proportion scored small values for one or two of these indicators. In most cases, weak indicators are compensated by excellent values obtained for the remaining one(s).

High values can also be obtained for very short or very long segments. These segments of improbable length should be discarded unless the configuration is very clear (when high values are obtained for all three indicators).

We introduce in the scoring formula a negative indicator, which performs a filtering of the probable transmembrane segments depending on their length. This is calculated with:

$$LP_l = e^{|l-21}$$

where $LP_i$ represents the length-penalty to be applied to a possible transmembrane segment of length *l*.

Each one of the four indicators should contribute with the same weight in the evaluation of the score for a segment. After normalisation of the hydrophobicity parameter, the score of a sequence from *m* to *p* is calculated by:

$$Score_m^p = e^{LeftInd_m} + e^{NH_m^p} + e^{RightInd_{p+1}} - Lp_l$$

where $l = p-m+1$ is the length of the sequence and $NH_m^p$ the average hydrophobicity for a segment of ten amino acids (normalised to a decapeptide) defined by $NH_m^p = \frac{10 H_m^p}{l}$ .

*Prediction algorithm*

For each position $m$ in the sequence, the maximum score that can be obtained if this position corresponds to the beginning of a transmembrane region is calculated:

$$MScore_m = \max\left( Score_m^p \right)$$

where $p$ varies from $m+1$ to $m+40$. It is ensured that the score is calculated for segments with positive indicators ( $LeftInd > 0$ and $RightInd > 0$ ). Concerning the hydrophobicity indicator, only the segments with $NH_m^p$ higher that a certain cut-off are kept (see Results).

Table I. Values obtained during the processing of the segment from residue 276 to residue 325 of the protein 5HT3_MOUSE (SwissProt protein code) utilising PRED-TMR.

| Pos | AA | $Mscore_m$ | End | TM |
|---|---|---|---|---|
| 276 | E | | | |
| 277 | R | | | |
| 278 | V | | | |
| 279 | S | | | |
| 280 | F | | | |
| 281 | K | | | |
| 282 | I | 34 | 303 | 2 |
| 283 | T | 11 | 303 | 2 |
| 284 | L | 19 | 304 | 2 |
| 285 | L | 10 | 304 | 2 |
| 286 | L | 3 | 308 | 2 |
| 287 | G | | | 2 |
| 288 | Y | 8 | 308 | 2 |
| 289 | S | 8 | 310 | 2 |
| 290 | V | 31 | 310 | 2 |
| 291 | F | 17 | 312 | 2 |
| 292 | L | 11 | 314 | 2 |
| 293 | I | | | 2 |
| 294 | I | | | 2 |
| 295 | V | | | 2 |
| 296 | S | | | 2 |
| 297 | D | | | 2 |
| 298 | T | 23 | 321 | 2 |
| 299 | L | 43 | 321 | 2 |
| 300 | P | | | 2 |
| 301 | A | 59 | 321 | 2 |
| 302 | T | | | 2 |
| 303 | I | 69 | 321 | 2 |
| 304 | G | | | |
| 305 | T | | | |
| 306 | P | 60 | 324 | |
| 307 | L | 89 | 324 | 1 |
| 308 | I | 74 | 330 | 1 |
| 309 | G | 70 | 330 | 1 |
| 310 | V | 80 | 330 | 1 |
| 311 | Y | 70 | 330 | 1 |
| 312 | F | 72 | 330 | 1 |
| 313 | V | 49 | 330 | 1 |
| 314 | V | 12 | 330 | 1 |
| 315 | C | | | 1 |
| 316 | M | | | 1 |
| 317 | A | | | 1 |
| 318 | L | | | 1 |
| 319 | L | | | 1 |
| 320 | V | | | 1 |
| 321 | I | | | 1 |
| 322 | S | | | 1 |
| 323 | L | | | 1 |
| 324 | A | | | 1 |
| 325 | E | | | |

Pos indicates the position in the sequence and AA shows the amino acid sequence itself (one-letter code). $MScore_m$ and End are the maximum score obtained and the corresponding end position for this score, respectively. The transmembrane segments detected are indicated in the TM column with a digit: 1 is used for the first segment found and 2 represents the second one. The observed (putative) transmembrane segments, as annotated in the SwissProt database, are shown in grey, for comparison.

For each position, the $MScore_m$ obtained and the corresponding end position are memorised. In the table generated, the highest $MScore_m$ is selected and the corresponding region is marked as transmembrane. Then, the second highest $Mscore_m$ is selected that does not overlap with a previously marked region and this process is continued with the next $Mscore_m$, until all possible regions are found.

As an example, consider the table of $MScore_m$ obtained for the segment from residue 276 to residue 325 of 5HT3_MOUSE (Table I). On this table, the program selects the highest $MScore_m$ (89 at position 307) and marks the segment from 307 to 324 as transmembrane. Then, it selects the second possible highest $Mscore_m$ . 80 at position 310 cannot be selected because this position is part of the first selected transmembrane domain. Also, 69 at position 303 cannot be selected because it represents a segment that ends at position 321, inside the transmembrane domain. The next possible $MScore_m$ is 34, at position 282, that represents a transmembrane segment from residue 282 to residue 303. As it is not possible to select a third segment, the program ends. For this region of the protein with observed (putative) transmembrane segments at 278-296 and 306-324, the algorithm detects two transmembrane domains at 282-303 and 307-324.

## Results

The predicted transmembrane domains were compared to the experimentally determined topologies calculating for each sequence:

- the percentage of residues predicted correctly (agreement factor), *Q*, defined by Chou & Fasman (1979),
- the correlation coefficient, *C*, (Fisher, 1958; Matthews, 1975),
- the ratio of segment matches, SM, defined by Cserzo et al. (1997).

We have optimised the hydrophobicity indicator cut-off on a sub-set of 64 proteins of the set used by Rost et al. (1995) (the sequences 2MLT, GLRA_RAT, GPLB_HUMAN, IGGB_STRSP and PT2M_ECOLI which were not found in the public databases were not used). The best results were obtained when segments with $NH_m^p$ < 2 were discarded. On the set of 64 proteins, an agreement factor of 88.24% was obtained, a correlation coefficient of 0.79 and a ratio of segment matches of 0.945.

In order to test the PRED-TMR algorithm, we have collected all available sequences used in three recent papers (Rost et al. 1995; Rost et al. 1996; Cserzo et al. 1997) and we have discarded those with more than 25% homology. The resulting set contains 101 non homologous transmembrane proteins in total. Details of the results obtained are not shown here, but they can be downloaded together with the list of the transmembrane segment assignments from http://o2.db.uoa.gr/PRED-TMR/Results/.

The results of the test on this set of 101 proteins gave an average *Q* of 88.83%, a *C* of 0.80 and a ratio of segment matches SM, equal to 0.954. One protein (1%) has a correlation coefficient < 0.4 and 10 have a C < 0.6 (10%). These scores are similar to those obtained by excluding the proteins used for the optimisation of the hydrophobicity indicator cut-off (*Q*=87.81%, *C*=0.78 and SM=0.943).

Table II shows the results produced applying PRED-TMR and five other prediction methods on the set of 101 proteins. Looking at the correlation coefficient, PRED-TMR was found to perform slightly better than the two best methods, PHDhtm and tmPRED, on this set. Concerning the agreement factor, PRED-TMR performs in a similar way as tmPRED and TOPPRED, whereas for the ratio of segment matches it is slightly worse than PHDhtm, which is best.

Table II. Comparison table of the average results obtained utilising PRED-TMR and 5 other prediction methods on a test set of 101 non homologous proteins.

| Method | *C* | *Q (%)* | *SM* |
|---|---|---|---|
| DAS | 0.71 | 87.83 | 0.823 |
| PHDhtm | 0.78 | 87.52 | 0.970 |
| TOPPRED | 0.72 | 88.85 | 0.881 |
| SOSUI | 0.71 | 86.56 | 0.917 |
| TmPRED | 0.75 | 89.31 | 0.895 |
| PRED-TMR | 0.80 | 88.83 | 0.954 |

$C$ is the correlation coefficient, $Q$ the agreement factor and SM the ratio of segment matches (see Results).

Despite the errors contained in SwissProt, it is thought that a comparison between predicted transmembrane regions and annotated ones, in the entire database, is worthwhile. It can serve as a common test set for algorithms detecting (predicting) transmembrane domains.

SwissProt, release 35, contains 9392 transmembrane sequences with a total of 40672 transmembrane regions. We have not discarded for the test transmembrane segments with uncertain endpoints as we have done to establish the statistics. The PRED-TMR algorithm applied to all proteins contained in the SwissProt database, produces slightly lower values for the $Q$ and $C$ scores and a rather larger decrease of the ratio of segment matches ($Q$=86.14, $C$=0.73, SM=0.889) relative to the test set of 101 proteins mentioned above. Of the 9392 proteins, 1710 (18%) have $C'$s< 0.6.

## Discussion

The PRED-TMR algorithm is a very simple and fast algorithm, it is available freely through the Internet and it does not require any additional information other than the protein sequence itself. It is comparable in terms of accuracy to most popular prediction methods.

Since PRED-TMR is a very fast algorithm and requires only information contained in a protein sequence alone, it is foreseen that its most potential use will be its application to ORF's (Open Reading Frames) predicted by the various genome projects, and especially those ORF's that correspond to proteins with unknown function. Aided by a pre-processing stage which could identify whether the sequence under study pertains to a membrane protein, it will be useful in the recognition of transmembrane domains. Such a pre-processing stage is well under way in our laboratory (Pasquier & Hamodrakas, In preparation): It is a neural network-based system which classifies proteins into four classes: fibrous(structural), globular, mixed (fibrous and globular) and membrane. The PRED-TMR algorithm has already been applied to the ORF's predicted from two genome projects and these results are currently being studied in detail.

PRED-TMR can certainly be improved by selecting carefully a representative and reliable set of transmembrane proteins to build the different tables. Ambiguities and errors in the existing databases impose limitations to its accuracy. When the statistical parameters used in the scoring formula were derived from the set of the 64 proteins, which were used to optimise the hydrophobicity cut-off, instead of calculating them from the entire SwissProt database, the accuracy scores decrease if the PRED-TMR algorithm is applied to sets larger than the original set of the 64 proteins. This is certainly due to the small reference set and reflects some special features of its sequences. However, it is believed that, the most promising way to improve the accuracy of prediction is to alter the scoring formula. Indeed, it was found that the length penalty used is not the most appropriate because it handicaps too harshly segments with a length outside the [17...25] range. Several other parameters can be added to the scoring formula like the positive inside rule defined by von Heijne (1992). However, we are convinced that this kind of algorithm will always be limited by the problem of using a strict cut-off to the hydrophobicity indicator. Fuzzy-logic seems to be a good technique to overcome this limitation by introducing some haziness in decision making.

A WWW server running the PRED-TMR algorithm is available at http://o2.db.uoa.gr/PRED-TMR/

## Acknowledgements

The authors gratefully acknowledge the support of the EEC-TMR "GENEQUIZ" grant ERBFMRXCT960019.